\documentclass[aps, pra, a4paper, 10pt, showpacs, twocolumn]{revtex4-1}
\usepackage{amsmath, amssymb, amsthm, bm, textcomp}
\usepackage[T1]{fontenc}
\usepackage[latin9]{inputenc}
\usepackage{graphicx,graphics}
\usepackage{geometry}
\usepackage{color}
\newcommand{\ket}[1]{\left|{#1}\right\rangle}

\newcommand{\be}{\begin{equation}}
\newcommand{\ee}{\end{equation}}
\newcommand{\eea}{\end{eqnarray}}
\newcommand{\bea}{\begin{eqnarray}}

\begin{document}

\title{Hybrid architecture for encoded measurement-based quantum computation}

\author{M. Zwerger$^1$, H. J. Briegel$^{1,2}$ and W.\ D\"ur$^1$}
\affiliation{$^1$ Institut f\"ur Theoretische Physik, Universit\"at
  Innsbruck, Technikerstr. 25, A-6020 Innsbruck,  Austria.\\
  $^2$Institut f\"ur Quantenoptik und Quanteninformation der \"Osterreichischen Akademie der Wissenschaften, Innsbruck, Austria}
\date{\today}

\begin{abstract}
We present a hybrid scheme for quantum computation that combines the modular structure of elementary building blocks used in the circuit model with the advantages of a measurement-based approach to quantum computation. We show how to construct optimal resource states of minimal size to implement elementary building blocks for encoded quantum computation in a measurement-based way, including states for error correction and encoded gates. The performance of the scheme is determined by the quality of the resource states, where within this error model we find a threshold of the order of 10\% local noise per particle for fault-tolerant quantum computation and quantum communication.
\end{abstract}

\maketitle


\textit{Introduction ---} Quantum computation holds the promise of solving certain problems much faster than any classical computer could. Several models for quantum computation have been put forward \cite{NiCh,Fa04,Ra01} that do not only offer conceptual insights into the scope and power of quantum computation, but also put different requirements and emphasis for an experimental realization. While the standard quantum gate-based approach makes use of a modular structure where coherent operations are chosen from a finite set of elementary gates and applied sequentially, measurement-based quantum computation is centered on the usage of a universal, highly entangled resource state that is processed by single qubit measurements only \cite{Ra01,Br09}. For certain set-ups, the preparation of specific entangled states will be easier than the coherent manipulation of arbitrary states, in particular if two-qubit gates requiring interactions between different systems are involved. However, large resource states need to be prepared and stored for universal measurement-based quantum computation. It is hence natural to consider hybrid schemes, in which elements of different computational schemes are combined into a computational architecture that unifies the advantage of the different approaches. Here we report on such a hybrid architecture, where elementary blocks and gate sequences are performed in a measurement-based way, i.e. by preparing specific resource states, and then combined and measured in a sequential fashion as in the circuit model.

The emphasis of our approach lies on (fault-tolerant) measurement-based realization of quantum error correction and the manipulation of encoded quantum information (see also \cite{Kn04}). In any realistic scenario, quantum information needs to be protected against noise and decoherence, which can be done by non local encoding and making use of quantum error correction \cite{NiCh,Go97,Ra12,Sho95,Ste96}. Here we demonstrate how the elementary building blocks of such encoded quantum computation - including encoding, syndrome-read out, decoding, but also realization of encoded quantum gates-  can be done optimally in a measurement-based fashion. To this aim, we construct optimal resource states of minimal size for all these tasks, for arbitrary Calderbank-Shor-Steane (CSS) \cite{Ste96} and stabilizer codes \cite{Go97}. All these resource states are stabilizer states, which can in principle be prepared probabilistically or pre-purified using known entanglement purification schemes \cite{Du03} to achieve a higher fidelity of the resource state and consequently of the performed operations. Such a probabilistic state preparation does not jeopardize the deterministic character of the overall computational scheme. What is more, we show how these basic elements can be combined in a straightforward way to obtain optimal resource states for a combination of the different tasks, including, e.g., code switchers or encoded gates combined with error syndrome readout, that is built-in error correction.
These measurement-based elements are then applied sequentially as in a gate-based approach, providing a modular and flexible structure. The coupling between the building blocks corresponds to simple Bell-measurements.

The performance of the scheme is determined by the quality of the resource states. We first show that for all computations that only consist of Clifford operations, more than 13.5\% of local depolarizing noise per particle is tolerable. In this sense our architecture provides a fault-tolerant Clifford quantum computer with a very high error threshold. In particular, we obtain a fault-tolerant quantum memory and code switcher, and a scheme for long-distance quantum communication. Using magic state distillation \cite{Bra05} at the logical level, we find that the the threshold of $13.5\%$ local noise per particle also applies to universal fault-tolerant quantum computation. Notice, however, that these numbers can not be directly compared to other fault-tolerance thresholds (e.g. \cite{Kn04,Ra12,Ra06}), as they are based on a different error model. We are considering a consistent error model within the measurement-based framework, where the main source of noise is due to imperfect preparation of resource states and imperfect measurements, which we model by single-qubit depolarizing noise. Notice that the preparation of resource states can in principle be done in various ways, not necessarily involving gates, so a reduction to the usually considered gate-based error models is not meaningful.

\textit{Background ---}
\label{Sec_background}
We start by reviewing quantum error correction and measurement-based quantum computation.
In quantum error correction, quantum information is encoded in a non local way into several physical systems such that arbitrary errors acting on one or several of the physical qubits can be corrected. In the standard (gate-based) approach, a quantum circuit ${\cal C}$ transforms an (unknown) single qubit state $\alpha |0\rangle + \beta |1\rangle$  to an encoded $M$-qubit state $\alpha |0_L\rangle + \beta |1_L\rangle$. The code is designed in such a way that each possible error operator --which is typically described by (tensor products) of Pauli operators corresponding to bit-flip, phase flip or both errors on one or several qubits respectively-- maps the logical subspace spanned by $\{|0_L\rangle,|1_L\rangle$ to a different orthogonal two-dimensional subspace. Clearly, the number of correctable errors is limited by the available dimension of the Hilbert space.
Error syndrome read-out takes place by projecting the system onto one of these two-dimensional error-subspaces. This is typically done with help of $m$ additional ancilla systems and a circuit ${\cal S}$ acting on the total $M+m$ qubits. The state of the ancillas is subsequently measured, thereby reading out the error syndrome and projecting the remaining system onto one of the two-dimensional error-subspaces. The measurement outcome also determines the correction operation needed to undo the error and return the system to the initial logical subspace. Notice that by this measurement,  a discretization of errors is enforced, i.e. a superposition of different error-states is projected (probabilistically) onto one of the possibilities, and can then be subsequently corrected. This is also the reason why it is sufficient to consider only elementary Pauli errors, as these operators form a basis for all Hermitian operators and hence any error can be written as a superposition of elementary Pauli errors. Decoding works in a similar way as encoding, where the inverse unitary ${\cal C}^{\dagger}$ is applied.

A special class of error correction codes of particular importance are so-called CSS codes \cite{Ste96}. They are stabilizer codes, that is, the codewords are stabilizer states, and their error-correction properties can be understood using the stabilizer formalism \cite{Go97}. For CSS error correction codes, encoding, decoding- and syndrome readout circuits are Clifford circuits, i.e., they can be implemented using only Clifford gates and Pauli measurements \cite{Go97}. This property will be important for us later, when we show how to realize the corresponding circuits in a measurement-based way.

In measurement-based quantum computation one starts with an entangled resource state. A quantum circuit is translated to a single-qubit measurement pattern on the resource state. There are several resource states which allow for universal quantum computation, e.g. the 2D cluster states \cite{Ra01b}. It is important to note that Clifford gates are implemented by Pauli measurements and can be done in the very first step of the measurement-based computation. Alternatively they can be taken into account even beforehand, entering into the design of smaller, task dependent resource state. A circuit which contains only Clifford gates and Pauli measurements, and has $n$ input and $m$ output qubits can be implemented on a $n+m$ qubit graph state \cite{Raussendorf2003}. Clifford circuits are crucial in entanglement purification and quantum error correction.

\textit{Elementary building blocks ---}
\label{blocks}
\label{Sec_elementary}
We will now show how to efficiently construct resource states of minimal size to perform certain gates or circuits in a measurement-based fashion.
There are two different ways to construct the required resource states. The first one starts with a sufficiently large 2D cluster state \cite{Ra01b} and the measurement pattern for the desired map. One can then apply the rules for Pauli measurements on graph states \cite{He04} and obtains the resource state of minimal size.

Alternatively one can construct the resource state via the Jamiolkowski isomorphism, which relates a completely positive map and a state \cite{Ja72}. It is simply the state which results from applying the map to N maximally entangled Bell pairs $\ket{\phi^+}=1/\sqrt{2} \left(\ket{00} + \ket{11} \right)$, where N is the number of qubits on which the map acts.

In both cases the calculations can be done efficiently using the stabilizer formalism \cite{Go97}. There exist programs to do so \cite{Aa04,An06}.

\begin{figure}[ht]
\begin{picture}(210,135)
\put(-35,-85)
{\includegraphics[width=10cm]{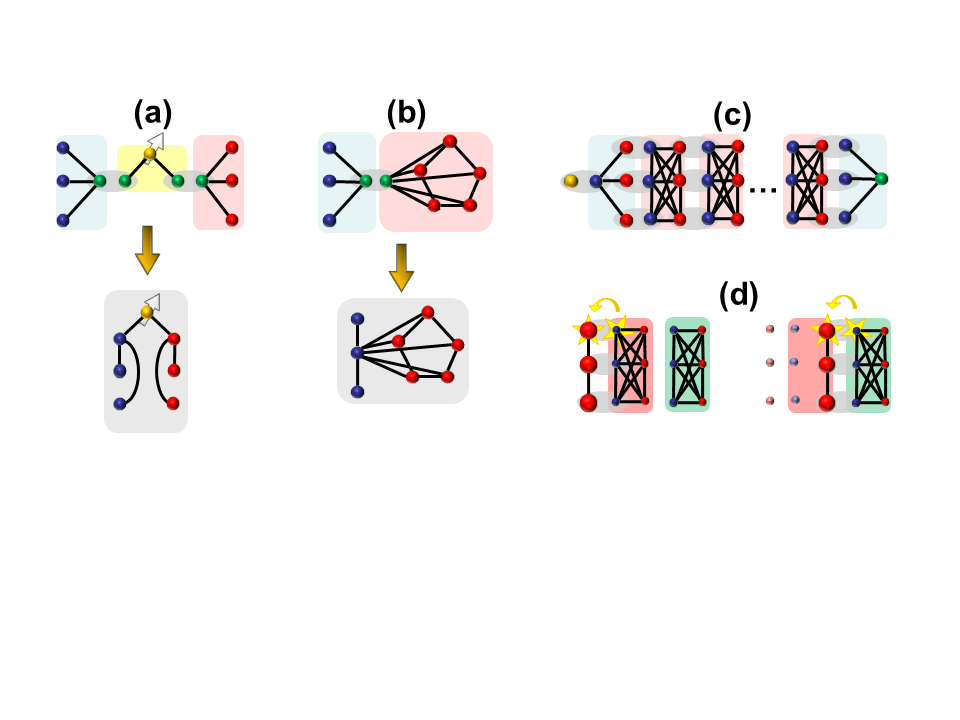}}
\end{picture}
\caption{(a) Illustration of fusion of resource states for a single qubit gate on a logical qubit. The resource states for decoding/encoding (left/right) are combined with the resource state for the rotation (middle) via Bell measurements. (b) Illustration of fusion of resource states for code switching. The resource states for decoding (left) and encoding (right) for two different codes are combined via a Bell measurement. (c) Resource states for encoding, repetitive error correction and decoding. (d) Two steps of the computation. Noise on input particles (blue) is moved to input particles (red). Noise on output particles is considered in the next step.}
\label{synthesis}
\end{figure}

A deterministic single qubit rotation around the $X$-axis can be implemented with the three qubit graph state shown in figure \ref{synthesis}a), yellow (middle) box. Here the qubit at the left hand side serves as input and the one on the right hand side as output. Also a two-qubit gate can be implemented deterministically using a 4-qubit resource state (see \cite{Sup}).
The CZ gate and the arbitrary single qubit gates form an universal set of gates, allowing to perform any quantum computation.

For all CSS codes, encoding and decoding require only Clifford gates and Pauli measurements. Particles that are measured in the Pauli basis can be removed from the resource state in a measurement-based implementation \cite{Zw12}. Consequently, such a circuit (independent of its length) can be done with a $M+1$ qubit resource state, where $M$ is the size of the code. The resource states for encoding and decoding are identical, since the decoding circuit is just the inverse of the encoding circuit. Resource states corresponding to a 3-qubit repetition code (capable of correcting a single-qubit bit-flip error) and a 5-qubit graph code (capable of correcting an arbitrary single qubit error) for encoding/decoding are shown in figure \ref{synthesis}b) (see \cite{Sup}).

The Bell measurements at the read-in of a logical qubit to a decoding resource state reveal the error syndrome, since the measurement results are correlated. Alternatively the syndrome could be obtained by constructing the resource state which implements the syndrome read-out (or syndrome read-out and decoding). In this construction it is assumed that the measurements of the ancilla qubits, which determine the syndrome, have a particular outcome. The byproducts from the Bell measurements will then lead to effective projections and reveal the error syndrome. This is similar to what has been used in measurement-based entanglement purification in order to determine whether a purification step was successful \cite{Zw12}.

The resource states described above can be combined to accomplish gates on the logical qubits. This is done in the following way: the output qubit of one resource state and the input qubit of the other resource state are measured in the Bell basis. This is illustrated for a single qubit rotation on a logical qubit and for code switching in figure \ref{synthesis}a,\ref{synthesis}b. In addition, the Bell measurements at the read-in reveal the error syndrome, so that error correction is combined with logical gates in a single step. The same technique can be used to design resource states for concatenated error correction or for combining logical gates, error correction and code switching. In all cases, one obtains resource states of minimal size.

\textit{Hybrid computational model ---}
\label{Sec_model}
We are now at a position to describe the hybrid architecture for encoded quantum computation. Processing of (encoded) quantum information takes place in a sequential way, where elementary building blocks are realized in a measurement-based fashion as described above. These elementary building blocks can correspond to rather complicated and lengthy quantum circuits, and error correction can always be included. The measurement-based realization allows one to implement them in a compact way using resource states of minimal size that only consist of input and output particles, independent of the length of the circuit. In addition, no ancilla particles are required, thereby reducing the number of involved particles. The complexity of the circuit is encoded in the specific structure of the resource state. The reduced size and complexity of this measurement-based implementation makes it less susceptible to noise, similar as encountered in measurement-based entanglement purification \cite{Zw13}.

Elementary building blocks are combined sequentially by performing Bell measurements between output particles of the previous block and input particles of the next block (see Fig. \ref{synthesis}c). The results of these measurement determine the error syndrome and the necessary correction operations. This modular structure limits the number of particles to be stored and operated on as compared to a purely measurement-based approach, as resource states can be prepared on the fly re-using the involved particles. In addition, no coherent manipulation by means of (entangling) gates is required, in contrast to the gate-based approach. Processing of (encoded) information is solely realized by preparation of small-scale resource states, together with Bell measurements. Notice that resource states may be prepared probabilistically and pre-purified using (multipartite) entanglement purification. As all resource states are graph states (up to local unitary operations), entanglement purification protocols are available \cite{Du03}. Such an approach may have significant advantages in certain set-ups such as for photons, especially since (probabilistic) creation of states is often a much easier task than coherent manipulation by means of gates.

\textit{Influence of noise and thresholds ---}
Imperfections in the resource state preparation, together with possibly imperfect Bell state measurements are the main sources of noise in this set-up.  As we are dealing with processing encoded information, we have to make sure that despite these effects and external decoherence, the noise at the logical level can be fully suppressed.

As a noise model, we consider local depolarizing noise (LDN) acting on each of the qubits individually,  ${\cal{D}}^j(p)\rho=p\rho + \frac{1-p}{2}\mathbb{I}^j\otimes \operatorname{tr}^j\rho$. Noisy resource states are thus given by $\rho_{R}={\cal{D}}(p)\rho=\left(\prod_{j=1}^{n}{\cal{D}}^j(p)\right)\rho$, where $n$ is the size of the resource state and $p$ specifies the amount of noise per particle. The effect of imperfect Bell measurements can be incorporated into local depolarizing noise for each particle, i.e. into the parameter $p$.
In addition, we describe the effect of decoherence and other imperfections by depolarizing noise with parameter $q$ acting on each of the particles of the encoded system.

As shown in \cite{Zw13}, LDN can be effectively moved to other particles when performing Bell measurements. Resource states for (encoded) Clifford gates and circuits (even with built-in error correction) only contain input and output particles. The key observation is that noise coming from imperfect preparation of resource states can be moved in part to the input state, and to the output state (after processing, i.e. it is considered in the next step of the computation). The total effect is that each particle of an encoded system is influenced by LDN twice coming from imperfect state preparation, plus noise coming from depolarization, i.e. ${\cal{D}}^j(p^2q)$, before it is processed {\em perfectly} in the desired way, i.e. by an encoded Clifford gate with built-in error correction (see Fig. \ref{synthesis}d). Since each of the encoded gates includes an error correction step, the noise at the logical level is reduced --corresponding to a larger value $q_L$- as long as $p^2q$ is above the threshold $p_{\rm Code}$ of the chosen error correcting code (w.r.t. depolarizing noise) \cite{Sm07,He05,Ke13}. Using the code in a concatenated fashion in fact ensures that $q_L \rightarrow 1$ in this regime, i.e. noise at the logical level can be fully suppressed \cite{He05,Ke13}. A sequential application of several of such logical operations is hence also possible \cite{fn1}.
As error correction can in principle be done very frequently, one can assume that $q \approx 1$. This is also justified by the fact that no active feed forward is required as long as only Clifford operations are involved. In fact, all correction operations can be postponed until the end of the logical circuit that may involve multiple blocks. It follows that the threshold $p=p_{\rm crit}$ for depolarizing noise per particle is simply given by $p_{\rm crit}=\sqrt{p_{\rm Code}}$. In a situation where $q=p$, one finds $\tilde p_{\rm crit}=q_{\rm crit}=\sqrt[3]{p_{\rm Code}}$. Thresholds for a concatenated 5-qubit code \cite{He05,Ke13} and Shor-type codes \cite{Sm07} are given by $p_{\rm Code}=0.8250$ [$p_{\rm Code}=0.7449$] respectively. This leads to $p_{\rm crit}=0.8631$ and $\tilde p_{\rm crit}=q_{\rm crit}=0.9065$ for the Shor-type codes.

{\em Applications ---}
As imperfections in resource states and Bell measurement are taken into account, we have that all Clifford circuits, including error correction circuits, codes switchers and logical Clifford gates for CSS codes, and even a quantum memory, can be implemented fault tolerantly in this way. The tolerable noise per particle for imperfect resource states is more than 13.5\%. One can also apply these results to long-distance quantum communication, where encoded quantum information is sent through noisy quantum channels and errors are repetitively corrected \cite{KnLa96}. As channel segments should not be too short, one has $q<1$. For $p=q$ one has $\tilde p_{\rm crit}=q_{\rm crit}=0.9065$, i.e. more than 9\% noise per particle for imperfect resource states and for channel noise is acceptable in this case.
This could provide a viable alternative for long distance quantum communication, with thresholds comparable to measurement-based quantum repeaters \cite{Zw12}.

Universal quantum computation also requires the implementation of non-Clifford gates. For single-qubit encoded non-Clifford gates, the yellow (upper) qubit in figure \ref{synthesis}a) has however to be treated differently, since it is not coupled to a logical qubit. Noise acting on it translates directly to an error on the logical qubit, and hence above arguments do not apply.
This can be avoided by using a fault-tolerant code switch (as described above) to an error correction code that allows for a transversal implementation of logical single qubit rotations, e.g. $\pi/8$ gates. Such a code was introduced in \cite{Bra05}, which however suffers from a relatively low error threshold of $p_{\rm code} \approx 0.981$ \cite{Ra06}. This leads to a threshold of about 1\% local noise per particle for resource states \cite{Sup}.
An alternative to above method, however with significant overheads, is given by using magic states and state injection \cite{Bra05} for implementing a single qubit, non Clifford rotation (see also \cite{Kn04}). Doing this again in a measurement-based fashion, we find that the threshold is the same as for Clifford computations, i.e. 13.5\% local noise per particle are acceptable \cite{Sup}.
Notice, however, that these numbers can not directly be compared to other threshold results for universal fault tolerant quantum computation, as they are based on a different error model.

\textit{Conclusions and outlook ---}
We have proposed a hybrid architecture for quantum computation, where elementary building blocks corresponding to error correction or encoded gates are implemented in a measurement-based fashion. This can be done with resource states of minimal size, thereby reducing the number of involved systems and the operational complexity. At the same time, we obtain very high error thresholds, where local depolarizing noise of the order of 10\% per particle is tolerable for the involved resource states. The crucial assumption in the derivation is that all noisy resource states can be described by local, uncorrelated noise acting on the individual qubits. Whether this can be justified for gate-based or other generation procedures remains an open issue. Our scheme is perfectly suited for several set-ups, and following our proposal, elementary building blocks corresponding to measurement-based quantum error correction have already been experimentally demonstrated with photons \cite{Ba13} and ions \cite{La13}.

\textit{Acknowledgements ---}
We thank D.E. Browne for helpful comments on the manuscript. This work was supported by the Austrian Science Fund (FWF): P24273-N16, SFB F40-FoQus F4012-N16.

\newpage

\newpage
\section*{Supplemental material}

\section{Graph states}

A graph $G$ is a set of vertices $V$ and edges $E$. In quantum information one associates a $N=|V|$ qubit graph state $\ket{G}$ with a graph $G$ in the following way: for each vertex $j$ one defines an operator $K_j=X_j \prod_{i \in N(j)}Z_i$, where $N(j)$ denotes the neighborhood of vertex $j$, i.e., all vertices that are connected to $j$ by an edge, and $X$ and $Z$ are the usual Pauli matrices. The graph state $\ket{G}$ is then uniquely defined as the eigenstate with eigenvalue $+1$ for all operators $K_j$. Alternatively $\ket{G}$ can be defined as the state resulting from the application a controlled phase gate ($CZ = diag(1,1,1,-1)$ in computational basis) between any pairs of qubits, initially in state $\ket{+}=1/\sqrt{2}\left(\ket{0}+\ket{1}\right)$, connected by an edge in the graph.
Graph states are important in the context of measurement-based quantum computation and quantum error correction.

\section{Quantum gates}

\subsection{Measurement-based implementation of single qubit gate}
Here we describe how a single-qubit rotation can be implemented in a measurement-based way using a 3-qubit graph state. The graph state is given by
\be
\ket{G_3}= \frac{1}{\sqrt{2}} \left( \ket{0}\ket{+}\ket{+} + \ket{1}\ket{-}\ket{-} \right).
\ee
The rotation angle $\alpha$ is realized via the choice of the measurement basis of the remaining qubit at the top (here qubit 1): it is measured in the basis $ \{e^{i\alpha}\ket{0}+e^{-i\alpha}\ket{1}, e^{i\alpha}\ket{0}-e^{-i\alpha}\ket{1} \}$. The measurement on the top qubit is done after the read-in and thus allows to adjust for a possible Pauli byproduct operator at the read-in. Rotations around the $Z$-axis can be obtained by applying a Hadamard operation on the two qubits at the bottom. Since any single qubit rotation can be decomposed into $X$- and $Z$-rotations this allows for the implementation of arbitrary rotations.

If the rotation is a single qubit Clifford gate the measurement at the top qubit will be a Pauli measurement and can be done beforehand, leading to a smaller resource state.

\subsection{Measurement-based implementation of two qubit gate}
A measurement-based implementation of the controlled phase (CZ) gate can be done with a state
\be
\ket{G_4}=(\ket{0000}+\ket{0011}+\ket{1100}+\ket{1111})/2
\ee
The read-in is done on qubits 1 and 3, the output qubits are qubits 2 and 4. Potential Pauli byproduct operators from the read-in can be commuted through the CZ gate, leading to new Pauli operators on the output qubits. Consequently the implementation is deterministic.

\section{Examples for codes}

\subsection{Repetition code}

Let us consider the code with codewords $\ket{0_L}=\ket{+}\ket{+}\ket{+}$ and $\ket{1_L}=\ket{-}\ket{-}\ket{-}$. It can correct a single qubit $Z$ error on any of the three physical qubits. The resource state which allows one to do the encoding into this code is given by
\be
\label{encoding}
\frac{1}{\sqrt 2}(|0\rangle|0_L\rangle + |1\rangle|1_L\rangle)
\ee
This state is up to local Clifford operations equivalent to the graph state shown in Fig. 1a (right).
Note that by choosing $\ket{0_L}=\ket{0}\ket{0}\ket{0}$ and $\ket{1_L}=\ket{1}\ket{1}\ket{1}$ one obtains a code which can protect against single qubit $X$ errors.

The decoding can be done with the same state, where qubits 2,3,4 are used as input, and qubit 1 as output (see Fig. 1a, left). The results of the Bell measurements also reveal the error syndrome, i.e. we actually have decoding with built-in error correction.

\subsection{Cluster-ring code}

The codes described above cannot protect against an arbitrary single-qubit error. This can be achieved with the optimal five-qubit code. The codewords correspond to graph states of a closed 5-qubit ring, with $|0_L\rangle$ [$|1_L\rangle$] being the $+1$ [$-1$] eigenstate with respect to all correlation operators $K_j=Z^{(j-1)}X^{(j)}Z^{(j+1)}$ (addition is understood modulo 5), $K_j|0_L\rangle = |0_L\rangle$, $K_j|1_L\rangle = -|1_L\rangle$ for all $j$ and $|1_L\rangle = Z^{\otimes 5}|0_L\rangle$. It is easy to see that any single-qubit Pauli error maps the logical subspace onto an orthorgonal error-subspace. The resource state which allows to do the encoding into this code is given by
\be
\frac{1}{\sqrt 2}(|0\rangle|0_L\rangle + |1\rangle|1_L\rangle).
\ee
This state is up to local clifford operations equivalent to the graph state shown in Fig. 1b (right).

Again, the same resource state allows one to perform the decoding with built-in error correction.

\section{Code switch}

A resource state capable of code switching can be obtained by combining a resource state for decoding for one code with a resource state for encoding for a different code via a Bell measurement. The Bell measurement is done beforehand, so that the state of the qubit is never decoded and quantum information remains protected. This is illustrated in Fig. 1b in the main paper for switching between the repetition and the ring cluster code. The Bell measurement changes the structure of the resulting graph state. Note that the measurement results at the read-in provide the error syndrome, and hence the required (Pauli) correction operation at the output particles. The required correction operation can be determined by considering the two virtual steps separately. The decoding with built-in error correction leads to a Pauli correction for the output particle of this (virtual) step. This correction operation is then mapped by the encoding circuit to a Pauli operation (possibly affecting many qubits) on the output particles of the second step.

\section{Fault-tolerance threshold using code switchers and transversal single-qubit gates}
Here we describe how error thresholds for universal fault-tolerant computation can be obtained using code switchers and transversal single-qubit gates. A single qubit non-Clifford gate is realized by switching to an error correction code such as the 15-qubit code CSS code related to classical Reed-Muller codes \cite{Bra05}. In this code, an encoded logical $\pi/8$-gate can be implemented transversally, i.e. by single qubit rotations. Together with logical Clifford gates, this constitutes a universal gate set. The code switch can be done fault-tolerantly as described in the main text. The single qubit rotation is assumed to be noisy, ${\cal E}_U\rho = {\cal{D}}^j(q_U)U\rho U^\dagger$, where noise can however be moved to the next error correction step, simply further decreasing the quality of the input particles, i.e. ${\cal{D}}^j(p^2 q\times q_U)$ \cite{fn2}. The fault tolerance threshold of the computation is again given by the condition $p^2 q\times q_U > p_{\rm Code}$, i.e. the threshold of the code against depolarizing noise. In many physical systems, such as e.g. trapped ions or superconducting qubits, single qubit rotation can be done with extraordinary accuracy \cite{Be08}, i.e. $q_U \approx 1$. Even when assuming $p=q_U$, we have $p_{\rm crit}=\sqrt[3]{p_{\rm Code}}$. For the 15 qubit CSS code, we have $p_{\rm Code} = 0.981$ \cite{Ra06}, i.e. universal fault tolerant quantum computation is possible with errors of more than 0.64\% errors per particle and single-qubit gate, and almost 1\% per particle when assuming high-fidelity single-qubit rotations.

\section{Fault-tolerance threshold for using magic state distillation and state injection}
Here we describe how to use magic state distillation and state injection, together with fault-tolerant Clifford circuits, to achieve universal quantum computation with high error thresholds. We use the magic state distillation procedure introduced in \cite{Bra05} at the logical level, i.e. we prepare several copies of single-qubit logical states by encoding them using an error encoding circuit. Magic distillation works as long as the fidelity of the initial (logical) state is above
$(1+1/\sqrt{2})/2\approx 0.854$, i.e. $p> 0.8047$ for local depolarizing noise \cite{Re05}.
If we encode a noisy physical qubit using a measurement-based encoding circuit, noise at the physical particle and the state for read in (e.g. first particle in Eq. \ref{encoding}) will directly translate to noise at the logical level, i.e. $p_{\rm crit}>0.8047$. Notice that a direct read-in via a single qubit measurement on the resource state is preferable to coupling an additional (noisy) particle via a Bell measurement. As long as noise at the logical level is below $p_{\rm crit}$, an (almost) perfect magic state can then be prepared at the logical level with help of magic state distillation.

The implementation of the magic state distillation as well as the logical single-qubit gate realized by state injection also only requires Clifford circuits \cite{Bra05}. In the latter case, the Clifford operations act on the (logical) magic state and the encoded quantum information. The threshold for noisy resource states to implement (logical) Clifford circuits fault-tolerantly are given in main text for different codes, and can be as high as 13.5\% per particle. Hence the threshold is determined by the applicability of fault-tolerant Clifford computation rather than the threshold for magic state distillation, leading to 13.5\% acceptable local noise per particle.


\begin{thebibliography}{99}
\bibitem{fn1}
Given that the distance of such a noisy logical operation to the perfect unitary is sufficiently small, say $\epsilon$, it follows that the sequential application of $M$ such operations has a distance smaller than $M \epsilon$ \cite{Gi04}. Notice that $\epsilon$ can be made exponentially small with only a polynomial overhead, e.g. by using a concatenated error correction code.
\bibitem{fn2}
The implementation of such single qubit unitary operations could also be done in a measurement-based way, however this is overly complicated.
\end{thebibliography}
\end{document}